# Bulk superconductivity induced by Se substitution in self-doped BiCh$_2$-based compound CeOBiS$_{2-x}$Se$_x$

Ryosuke Kiyama[1], Yosuke Goto[1], Kazuhisa Hoshi[1], Rajveer Jha[1], Akira Miura[2], Chikako Moriyoshi[3], Yoshihiro Kuroiwa[3], Tatsuma D. Matsuda[1], Yuji Aoki[1], Yoshikazu Mizuguchi[1]

1. Department of Physics, Tokyo Metropolitan University, 1-1, Minami-osawa, Hachioji, 192-0397, Japan.
2. Faculty of Engineering, Hokkaido University, Kita-13, Nishi-8, Kita-ku, 060-8628, Japan
3. Department of Physical Science, Hiroshima University, 1-3-1, Kagamiyama, Higashihiroshima, 739-8526, Japan

**Abstract**

We report the Se substitution effects on the crystal structure, superconducting properties, and valence states of self-doped BiCh$_2$-based compound CeOBiS$_{2-x}$Se$_x$. Polycrystalline CeOBiS$_{2-x}$Se$_x$ samples with $x$ = 0–1.0 were synthesized. For $x$ = 0.4 and 0.6, bulk superconducting transitions with a large shielding volume fraction were observed in magnetic susceptibility measurements; the highest transition temperature ($T_c$) was 3.0 K for $x$ = 0.6. A superconductivity phase diagram of CeOBiS$_{2-x}$Se$_x$ was established based on $T_c$ estimated from the electrical resistivity and magnetization measurements. The emergence of superconductivity in CeOBiS$_{2-x}$Se$_x$ was explained with two essential parameters of in-plane chemical pressure and carrier concentration, which systematically changed with increasing Se concentration.

# 1. Introduction

Since the discovery of BiS$_2$-based superconductors Bi$_4$O$_4$S$_3$ and LaO$_{1-x}$F$_x$BiS$_2$, the BiCh$_2$-based (Ch: chalcogen) systems have been drawing much attention as a new class of layered superconductors [1-3]. The crystal structure is composed of alternate stacks of electrically conductive BiCh$_2$ layers and insulating (blocking) layers, which is similar to those of the Cuprate and the FeAs-based superconductors with a high transition temperature ($T_c$) [4,5]. Several types of BiCh$_2$-based superconductors have been synthesized and the structural and compositional variations have been widely developed. So far, many experimental and theoretical studies have been performed to elucidate the superconducting mechanisms of the BiCh$_2$-based compounds and to further increase $T_c$, but the conclusion about the superconductivity mechanisms has not been obtained [6,7]. In the early studies including theoretical calculation [8], Raman scattering [9], muon-spin spectroscopy [10], and thermal conductivity [11] experiments suggested conventional mechanisms with fully gapped s-wave. On the other hand, recent first principles calculation [12], angle-resolved photoemission spectroscopy [13], Se isotope effect [14], and in-plane anisotropy measurements [15] proposed unconventional pairing mechanism in BiCh$_2$-based superconductors.

The parent phase of the BiCh$_2$-based superconductor is a band insulator [1,3,6,7,16]. Based on the calculated band structure, carrier doping to conduction bands which are mainly composed of Bi-6p$_x$ and Bi-6p$_y$ orbitals is expected to induce metallicity. Actually, partial F substitutions for the O site in REO$_{1-x}$F$_x$BiS$_2$ (RE: rare earth or Bi) generate electron carriers [2,3]. Indeed, superconductivity has been observed in carrier-doped compounds, but, for some of the synthesized samples, filamentary superconductivity with a small shielding volume fraction in magnetic susceptibility measurements has been observed [1-3]. In BiCh$_2$-based compounds, the emergence of bulk superconductivity requires optimization of the local crystal structure as well as carrier doping. External pressure effects [17-20] and chemical pressure effects [21-24] are powerful methods to induce bulk superconductivity. Particularly, in-plane chemical pressure is an important factor for superconductivity in BiCh$_2$-based compounds [22]. One of the typical routes to increase in-plane chemical pressure is partial substitution of Se for the S site [23-26]. In LaO$_{0.5}$F$_{0.5}$BiS$_{2-x}$Se$_x$, since La-(O/F) bond length is not largely affected by Se substitution, an increase of amount of Se$^{2-}$ with larger ironic radius leads to enhancing in-plane chemical pressure [22]. As a result, metallicity and bulk superconductivity have been induced by the in-plane chemical pressure effect in LaO$_{0.5}$F$_{0.5}$BiS$_{2-x}$Se$_x$.

The target material of this study is CeOBiS$_2$ [27]. CeOBiS$_2$ is a typical REOBiCh$_2$-type material. F-substituted CeO$_{1-x}$F$_x$BiS$_2$ shows bulk superconductivity when the samples were annealed under high pressure [28]. In CeO$_{1-x}$F$_x$BiS$_2$, superconductivity and ferromagnetism coexist [27-29]. Furthermore, CeOBiS$_2$ has been investigated as a material located at a quantum critical point [30]. Notably, the parent phase CeOBiS$_2$ shows metal-like conductivity and superconductivity at 1.3 K without carrier doping, and the $T_c$ increases under high pressure [31,32]. This is inconsistent with the band calculations mentioned above. However, this metallicity in F-free CeOBiS$_2$ is most likely due to electron doping via mixed valence states of Ce ions [33]. Similar behavior has been observed in EuFBiS$_2$ and Eu$_3$Bi$_2$S$_4$F$_4$ [34-36]. We consider that self-doped BiCh$_2$-based superconductor is a promising system for elucidating mechanisms of superconductivity because of less disorder in the blocking layers.

In this study, we synthesized a BiCh$_2$-based superconductor system CeOBiS$_{2-x}$Se$_x$ ($x$ = 0–1.0) and investigated the crystal structural, transport, and magnetic properties. Metallic conductivity and bulk superconductivity were induced by Se substitution for $x$ = 0.4 and 0.6.

## 2. Experimental

Polycrystalline samples of CeOBiS$_{2-x}$Se$_x$ with $x$ = 0, 0.2, 0.4, 0.6, 0.8, and 1.0 were prepared by a solid-state reaction method. Powders of Bi$_2$S$_3$ and Bi$_2$Se$_3$ were pre-synthesized by sintering a mixture of grains of Bi (99.999%), S (99.99%), and Se (99.999%) in an evacuated quartz tube at 700 ºC for 15 h. Powders of Ce$_2$S$_3$ (99.9%), CeO$_2$ (99.99%), Bi$_2$O$_3$ (99.999%), Bi$_2$S$_3$, Bi$_2$Se$_3$ and Bi (99.999%) grains with a nominal composition of the target phases ($x$ = 0–1.0) were mixed, pressed into a pellet, sealed into an evacuated quartz tube, and heated at 700 ºC for 5 h. The obtained sample was ground, mixed, pelletized, and heated for 5 h at different temperature. Samples heated at 400 ºC were labelled S1 and samples heated at 700 ºC were labelled S2. S1 samples contained fewer impurities than S2 samples, but it was too fragile to measure electrical resistivity with four proves. Therefore, S1 was used for structural analysis, and S2 was used for electrical resistivity and magnetization measurements.

The obtained samples were characterized by X-ray diffraction method (XRD) with a Cu K$_\alpha$ radiation (Rigaku, Miniflex-600) by the $\theta$-$2\theta$ method. Synchrotron XRD experiments with λ = 0.495813(1) Å were performed using BL02B2 beamline at SPring-8 under the proposal of No. 2019A1101. The experiment was performed with Multiple MYTHEN system [37] with a step of $2\theta$ = 0.006º. A crystal structure image was depicted using a VESTA software [38]. Structure parameters were obtained from Rietveld analysis using RIETAN-FP [39]. Electrical resistivity measurements were performed using the four-terminal method. For $x$ = 0.2 and 1.0, the resistivity measurements were performed with a $^3$He system of Physical Property Measurement System (Quantum Design, PPMS) down to $T$ = 0.5 K. The temperature dependence of magnetic susceptibility was measured by a superconducting interface device (SQUID) magnetometer with an applied field 10 Oe after both zero-field cooling (ZFC) and field cooling (FC) using Magnetic Property Measurement System (Quantum Design, MPMS-3).

## 3. Results and Discussion

Figure 1(a) shows the XRD patterns for S2 samples ($x$ = 0.0–1.0). The major phase can be indexed with the CeOBiS$_2$-type tetragonal phase with the space group of $P4/nmm$, although minor diffraction peaks corresponding CeO$_2$ impurity (< 5 wt%) were also observed. As shown in Fig. 1(b), XRD peaks systematically shift according to the Se concentration. Figure 2 shows the XRD patterns and the Rietveld fitting result for S1 sample ($x$ = 0.6 as a representative sample). The small amount of CeO$_2$ impurity phase (< 2 wt%) was detected. XRD data in Figs. 1 and 2 show that the S1 and S2 samples are comparable except for the amount of impurities. Synchrotron XRD data was used for crystal structure analysis (Figs. 3 and 4).

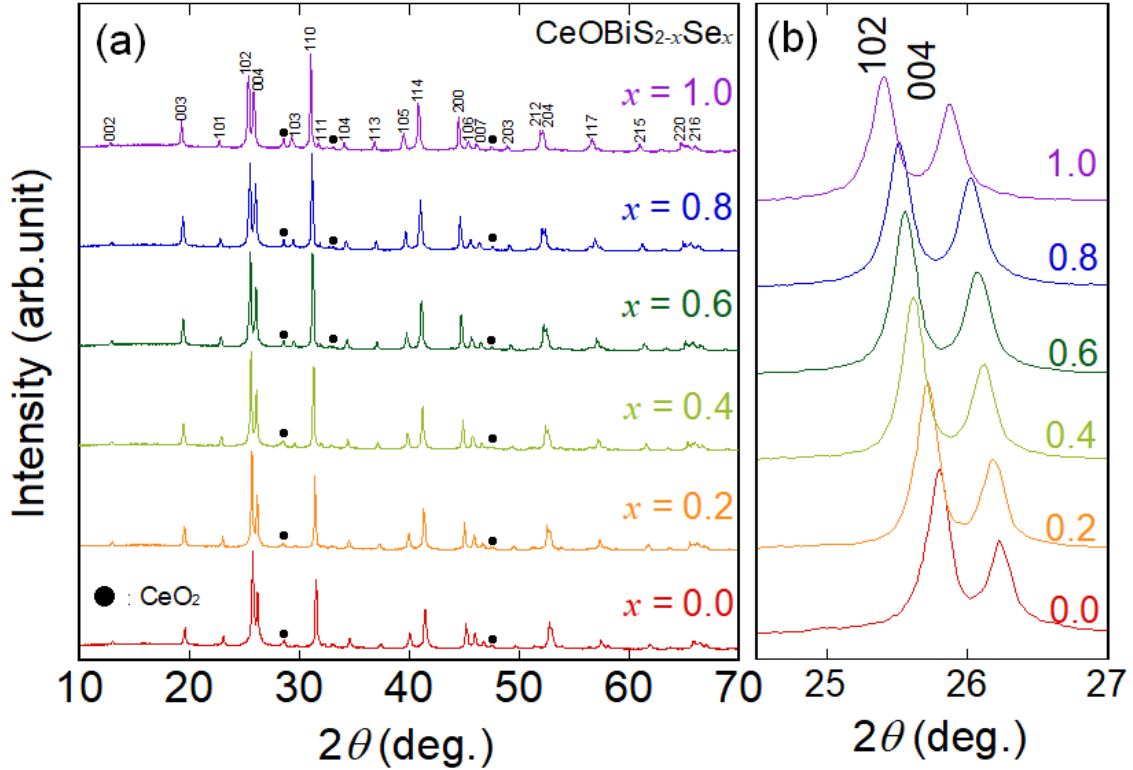

**Figure 1: (a) XRD patterns (Cu-K$_\alpha$) for CeOBiS$_{2-x}$Se$_x$ with $x$ = 0.0–1.0. Circles (●) indicate peaks resulting from a CeO$_2$ impurity phase. The numbers in the figure indicate Miller indices. (b) Zoomed profiles near 102 and 004 peaks.**

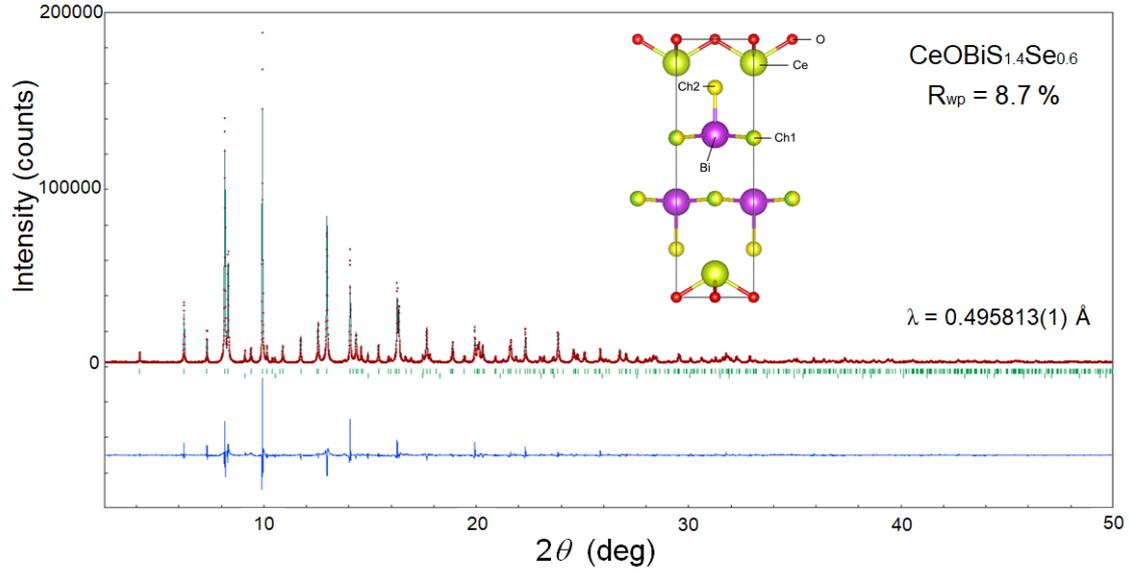

Figure 2: Synchrotron X-ray ($\lambda$ = 0.495813(1) Å) powder diffraction pattern and Rietveld analysis results for CeOBiS$_{1.4}$Se$_{0.6}$. The refinements were performed by two-phase analysis with a secondary phase of CeO$_2$. The blue profile plotted at the bottom shows a residual curve between observed and calculated data. The inset shows the schematic representation of crystal structure for CeOBiS$_{1.4}$Se$_{0.6}$. Ch1 and Ch2 denote the in-plane and out-of-plane chalcogen sites, respectively.

Figure 3(a) shows the $x$ dependences of the lattice parameters ($a$ and $c$) and the lattice volume. Both the lattice parameters increase with increasing $x$, which can be understood by the difference in the ionic radius of Se$^{2-}$ (198 pm, assuming a coordination number of 6) and S$^{2-}$ (184 pm, assuming a coordination number of 6). These elongations of the lattice parameters by Se substitution resemble those observed in LaOBiS$_{2-x}$Se$_x$ [40]. While Se ions dominantly occupy the in-plane Ch1 site, S ions occupy the out-of-plane Ch2 site. Thus, Se ions selectively occupy the in-plane Ch1 site, which is consistent with the observations in Se-doped REOBiCh$_2$-type compounds [26,40,41]. At $x$ = 1.0, approximately 85% of Ch1 and 15% of Ch2 sites are occupied with Se, respectively. Figure 3(c) shows interatomic distances as a function of $x$. The Bi-Ch1 (inter-plane), Ce-Ch2, and Bi-Ch1 (in-plane) distances increase with increasing $x$. The maximum change of interatomic distances by Se substitution ($\Delta R_{max}$) is about 0.07 Å, 0.05 Å and 0.04 Å for those distances, respectively. The Ce-O distance is almost constant but slightly increases ($\Delta R_{max}$ ~ 0.008 Å). In contrast, the Bi-Ch2 distance decreases ($\Delta R_{max}$ ~ −0.03 Å). The position of Ch2 with respect to the BiS$_2$ layer plays an important role in the doping mechanism in EuFBiS$_2$ [35]. Thus, the change in the Bi-Ch2 distance may be one of the important factors in this system as well.

In order to compare the amplitude of in-plane chemical pressure (CP) and physical properties in CeOBiS$_{2-x}$Se$_x$, we have estimated in-plane chemical pressure amplitude parameter, which is defined as CP = (R$_{Bi}$ + R$_{Ch1}$) / (Bi-Ch1 in-plane distance), where R$_{Bi}$ and R$_{Ch1}$ are the ionic radii of Bi and Ch1, respectively [22]. We assume that the valence of Bi is +3 and the R$_{Bi}$ is 1.03 Å. R$_{Ch1}$ is 1.84 and 1.98 Å for S$^{2-}$ and Se$^{2-}$, respectively. We have estimated the Bi-Ch1 (in-plane) distance using the structural parameters obtained from Rietveld refinement (Fig. 3(c)). Figure 3(d) shows the $x$ dependence of the in-plane CP for CeOBiS$_{2-x}$Se$_x$. The CP increases with increasing $x$, and this trend of in-plane CP in CeOBiS$_{2-x}$Se$_x$ is almost the same as that observed in La(O,F)BiS$_{2-x}$Se$_x$ and Eu$_{0.5}$La$_{0.5}$FBiS$_{2-x}$Se$_x$ [22, 41,42].

Figure 3(e) shows the bond valence sum for the Ce site (Ce-BVS) plotted as a function of $x$. As demonstrated in Ref. 43 the Ce-BVS was calculated using the following parameters: $b_0$ = 0.37 Å, $R_0$ = 2.151 Å for Ce-O bond, 2.62 Å for Ce-S bond and 2.74 Å for Ce-Se bond. Bond distances between Ce and nine coordinating anions were determined by Rietveld refinement. Site occupancies at the chalcogen site were included in the Ce-BVS calculation. The Ce-BVS tends to decrease with increasing $x$. This result suggests the mixed valence of Ce in CeOBiS$_{2-x}$Se$_x$. Qualitatively, in order to maintain electrical neutrality, the valence of the conductive layer should increase as the valence of the blocking layer decreases. Thus, the decrease of Ce valence implies reduction of the number of self-doped electrons in the conductive layer.

Figures 4(a) and (b) show $x$ dependences of $U_{11}$ and $U_{33}$ for both the Bi and Ch1 sites. For the anisotropic analysis of the displacement parameter $U$, the in-plane $U_{11}$ and the out-of-plane $U_{33}$ were refined by Rietveld refinement. The $U_{11}$ for the Ch1 site is clearly larger than that for Bi site for $x$ = 0 and 0.2, and it decreases with increasing $x$. The origin of the large $U_{11}$ for Ch1 has been suggested to be the presence of local in-plane disorder from extended X-ray absorption fine structure, X-ray diffraction, and neutron diffraction for several BiCh$_2$-based system [3, 42, 44-50]. The reduction of large $U_{11}$ for Ch1 indicates that the in-plane local disorder was suppressed due to the in-plane CP. Recent studies have shown that the increase in the in-plane displacement of Ch1 would induce carrier localization and be negatively link to superconductivity [3]. Therefore, Se substituted samples with $x$ > 0.2, which have less in-plane disorder, are expected to show bulk superconductivity in the CeOBiS$_{2-x}$Se$_x$ from structural point of view. On the $U_{33}$ for Bi and Ch1 sites, $U_{33}$ increases by Se substitution for both sites as displayed in Fig. 4(b). This increase in $U_{33}$ is probably related to the huge atomic vibration along the inter-layer direction which has been proposed for LaOBiS$_{2-x}$Se$_x$. The anharmonic vibration of atoms in the in-plane site is enhanced by Se substitution in LaOBiS$_{2-x}$Se$_x$ [51].

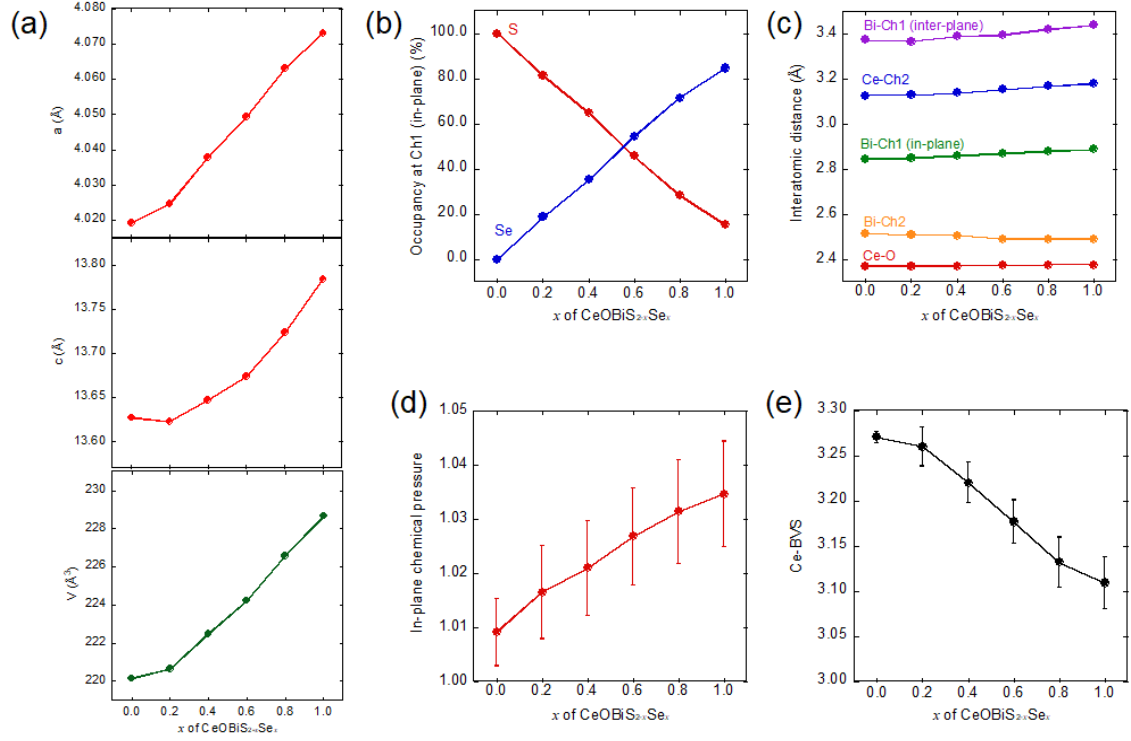

**Figure 3: (a)** Lattice parameters (*a* and *c*) and lattice volume (*V*) of CeOBiS$_{2-x}$Se$_x$. **(b)** S and Se occupancy at the in-plane Ch1 site in CeOBiS$_{2-x}$Se$_x$ as a function of *x*. **(c)** Interatomic distances in CeOBiS$_{2-x}$Se$_x$ as a function of *x*. **(d)** In-plane chemical pressure amplitude for CeOBiS$_{2-x}$Se$_x$ as a function of *x*. **(e)** Bond valence sum for Ce (Ce-BVS) as a function of *x*.

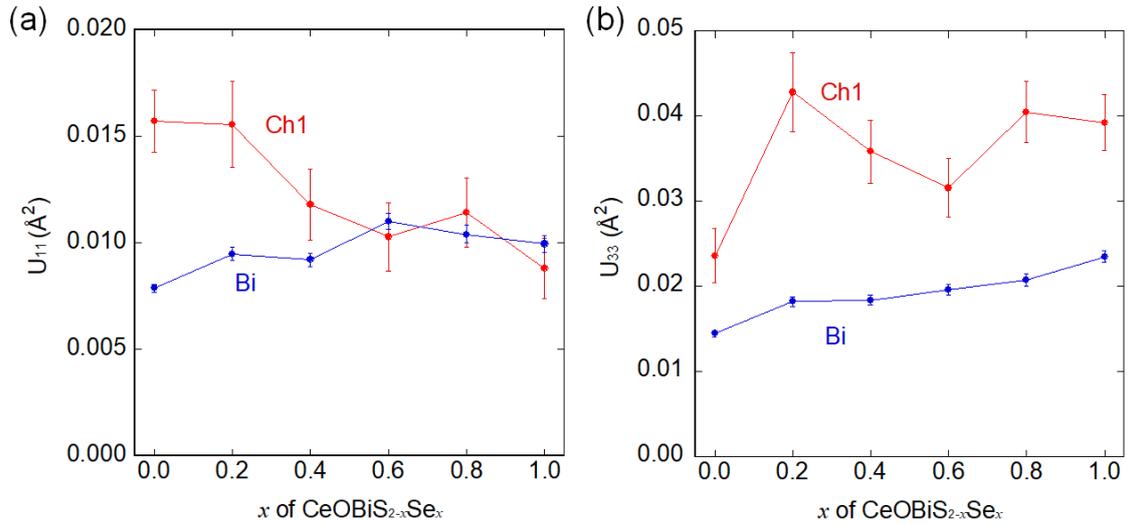

**Figure 4:** Se concentration dependences of the anisotropic displacement parameters **(a)** $U_{11}$ and **(b)** $U_{33}$ for the Bi and Ch1 sites in CeOBiS$_{2-x}$Se$_x$.

Figures 5(a) and (b) show the temperature dependences of electrical resistivity for $x$ = 0.0–1.0. For $x$ = 0, the temperature dependence of resistivity shows slight increase with decreasing temperature, which is different from the previous work [27]. This difference in transport properties may be due to the local in-plane disorder. Furthermore, an anomaly was observed at 90–180 K in the parent phase, in a similar manner as that observed in a previous work [28]. With increasing $x$, resistivity decreased, and metallic conductivity was observed for $x$ = 0.6–1.0. The enhanced metallicity in Se-substituted samples should be due to the increase in in-plane chemical pressure. A superconducting transition was observed for $x$ = 0.2-0.8. The highest onset temperature ($T_c^{onset}$) and the highest zero-resistivity temperature ($T_c^{zero}$) were 3.0 K ($x$ = 0.6) and 2.7 K ($x$ = 0.4 and 0.6), respectively. Notably, the anomaly observed for $x$ = 0 was suppressed in $x$ = 0.2 and completely disappeared in $x$ = 0.4. This result implies that the parent phase may have charge-density-wave (CDW)-like instability, which has been also proposed in EuFBiS$_2$ [34]. In order to determine the cause of this anomaly, detailed crystal structure analysis using single crystals at low temperature may be needed.

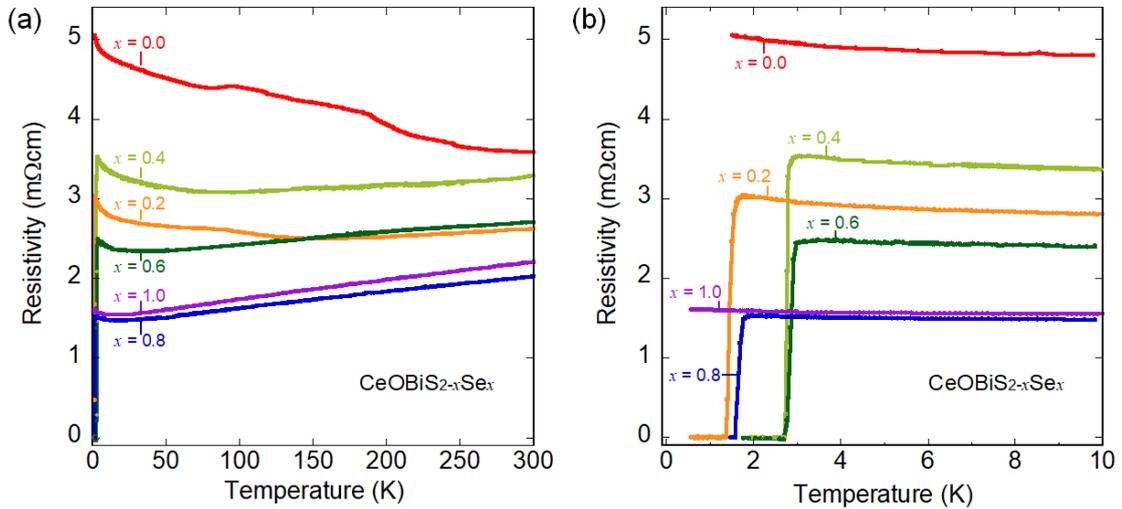

**Figure 5: (a) Temperature dependences of electrical resistivity for CeOBiS$_{2-x}$Se$_x$. (b) Enlarged picture of electrical resistivity below 10 K.**

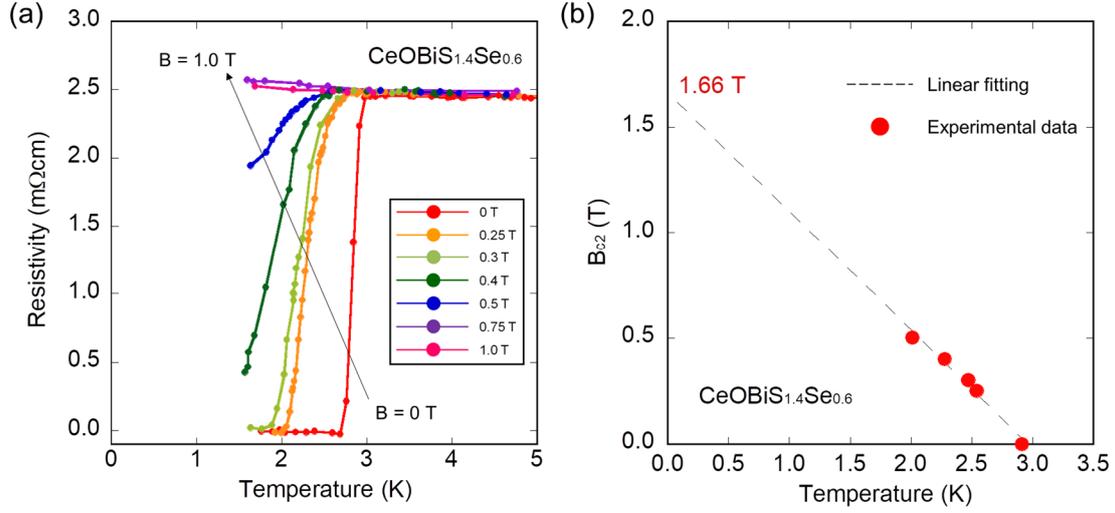

**Figure 6: (a) Temperature dependences of resistivity for CeOBiS$_{1.4}$Se$_{0.6}$ under magnetic fields up to 1 T. (b) of $B_{c2}$ vs. $T$ phase diagram for CeOBiS$_{1.4}$Se$_{0.6}$.**

Figure 6(a) shows the temperatures dependence of electrical resistivity under various magnetic fields for CeOBiS$_{1.4}$Se$_{0.6}$, which showed the highest $T_c$. $T_c$ shifts towards the lower temperature side with increasing magnetic field. An upper critical field ($B_{c2}$)-temperature phase diagram is shown in Fig. 6(b). $B_{c2}$ are estimated at temperatures where the resistivity reaches 90% of normal-state values under various applied magnetic fields. We calculated $B_{c2}(0)$ by the conventional one-band Werthamer-Helfand-Hohenberg (WHH) model [52], which gives $B_{c2}(0) = -0.693T_c(dB_{c2}/dT)_{T\ T_c}$. $B_{c2}(0)$ obtained from the WHH method is 1.15 T.

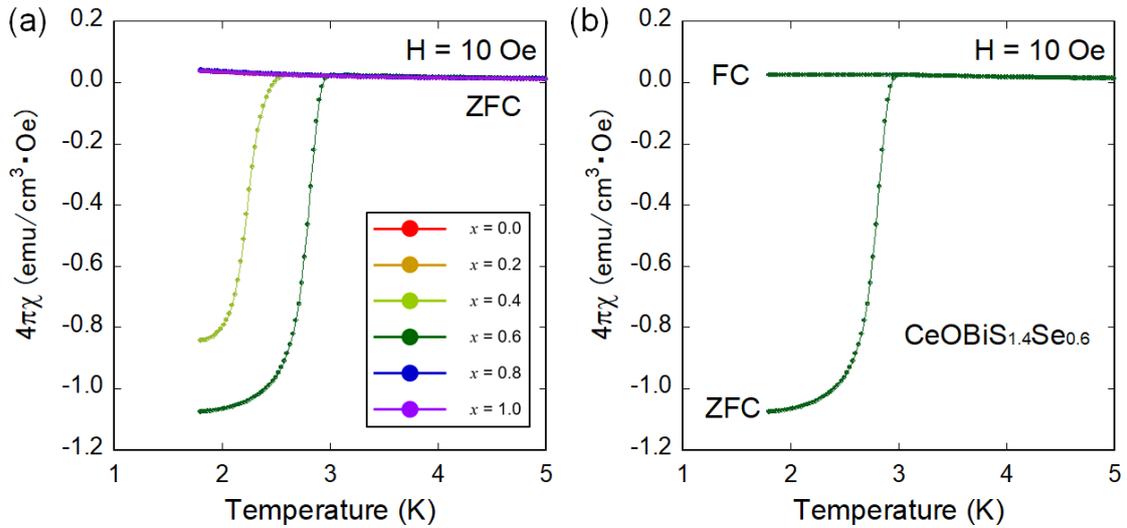

**Figure 7: (a) Temperature dependences of magnetic susceptibilities ($4\pi\chi$-ZFC) for CeOBiS$_{2-x}$Se$_x$ ($x$ = 0.0–1.0). (b) ZFC and FC data for $x$ = 0.6.**

Figure 7 shows the temperature dependences of magnetic susceptibility below 5 K with an applied magnetic field of 10 Oe after ZFC for CeOBiS$_{2-x}$Se$_x$. A large diamagnetic signal corresponding to superconductivity were observed for $x$ = 0.4 and 0.6, indicating that the observed superconducting states are bulk in nature. For the samples with $x$ = 0.2 and 0.8, no diamagnetic signal was observed down to 1.8 K, while those samples showed zero resistivity in the electrical resistivity measurement (Fig. 5(b)). Although measuring specific heat around $T_c$ is one of the best ways to check the bulk nature of superconductivity, it is challenging to characterize a superconducting transition from specific heat because magnetic entropy of Ce 4f electrons should obscure the electronic specific heat jump at a superconducting transition [30]. The largest shielding volume fraction and the highest $T_c$ of 3.0 K were obtained for $x$ = 0.6, which is slightly higher than the one reported for CeO$_{0.5}$F$_{0.5}$BiS$_{2-x}$Se$_x$ [53]. Regarding possible coexistence of ferromagnetism and superconductivity observed in the CeO$_{1-x}$F$_x$BiS$_2$ system [27-29], our present results did not show any coexistence of superconductivity and magnetic ordering.

On the basis of the measured results of electrical resistivity and magnetization, a superconductivity phase diagram was established (Fig. 8); the $T_c$ data for $x$ = 0 was taken from Ref. 31 and plotted as well. Our results show that Se substitution suppresses carrier localization, described as "bad metal" in the phase diagram, and induces bulk superconductivity; this trend can be understood by in-plane chemical pressure effects. Then, $T_c$ decreases and superconductivity phase disappears at higher Se concentrations. The suppression of superconductivity is probably linked to the decreased carrier concentration due to the decrease in Ce valence by Se substitution. From this work, we have obtained a bulk superconducting sample with less disorder in the blocking layer such as CeOBiS$_{1.4}$Se$_{0.6}$, which will be useful for experiments to probe mechanisms of superconductivity in BiCh$_2$-based compounds.

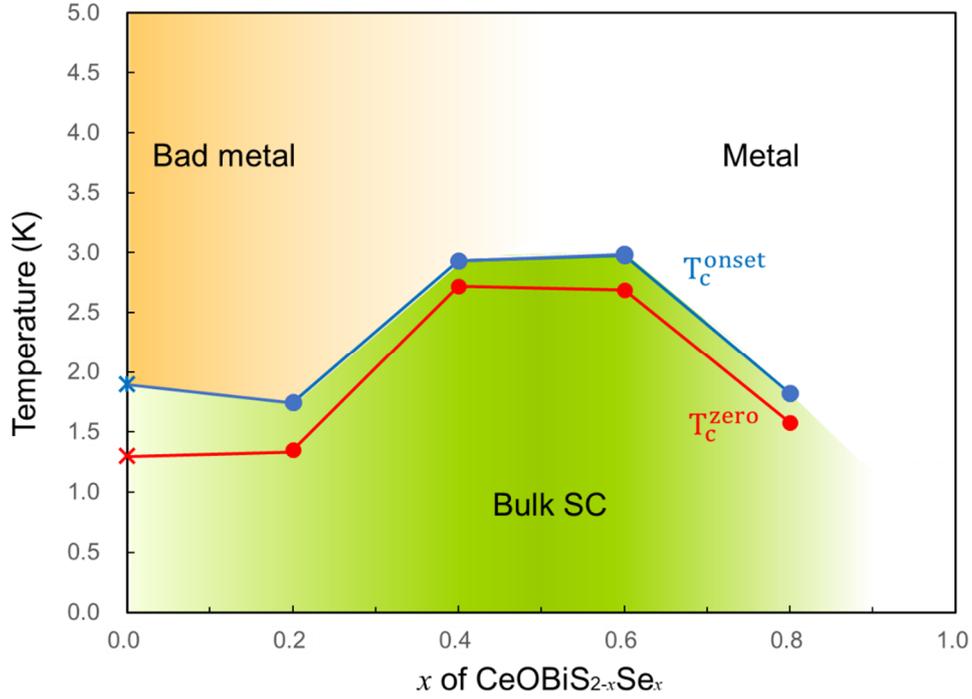

**Figure 8:** Superconductivity phase diagram of CeOBiS$_{2-x}$Se$_x$. The data for $x = 0$ was taken from Ref. 31.

## 4. Conclusions

We have synthesized a new BiCh$_2$-based superconductor system CeOBiS$_{2-x}$Se$_x$, which shows a bulk superconductivity when local structural parameters and carrier concentration were both optimized by Se substitution. Polycrystalline samples of CeOBiS$_{2-x}$Se$_x$ with $x = 0$–1.0 were synthesized by the solid-state reaction method. The XRD and Rietveld analyses revealed that in-plane chemical pressure was enhanced, and the in-plane disorder was suppressed by Se substitution. BVS calculation showed that the valence of Ce ions decreased by Se substitution. For $x = 0.4$ and 0.6, a superconducting transition with a large shielding volume fraction was observed in magnetic susceptibility measurements, and the highest $T_c$ was 3 K for $x = 0.6$. From electrical resistivity measurements, a zero-resistivity state was observed for $x = 0.2$-0.8, and the highest $T_c$ was observed for $x = 0.6$. Since the CeOBiS$_{1.4}$Se$_{0.6}$ superconductor has less disorder in the CeO blocking layer owing to no substitution, this superconducting material should be useful for experiments to probe mechanisms of superconductivity in BiCh$_2$-based compounds.

**Acknowledgements**

We thank O. Miura and N. L. Saini for their support on experiments and discussion. This work was partly supported by grants in Aid for Scientific Research (KAKENHI) (Grant Nos. 15H05886, 15H05884, 16H04493, 18KK0076, and 15H03693) and the Advanced Research Program under the Human Resources Funds of Tokyo (Grant Number: H31-1).